\documentclass[conference,10pt]{IEEEtran}
\IEEEoverridecommandlockouts

\usepackage{epsfig,rotating,setspace,latexsym,amsmath,epsf,amssymb,amsfonts,bm,theorem,cite,algorithm,graphicx,epsf,authblk,epstopdf,color,algpseudocode,bbm}

\usepackage[algo2e]{algorithm2e}

\newtheorem{remark}{Remark}

\newtheorem{lemma}{Lemma}
\newenvironment{Proof}[1]{\medskip\par\noindent{\bf Proof:\,}\,#1}{{\mbox{\,$\blacksquare$}\par}}

\allowdisplaybreaks

\begin{document}

\title{Timely Transmissions Using Optimized\\Variable Length Coding
}

\author[1]{Ahmed Arafa}
\author[2]{Richard D. Wesel}
\affil[1]{\normalsize Electrical and Computer Engineering Department, University of North Carolina at Charlotte, NC 28223}
\affil[2]{\normalsize Department of Electrical and Computer Engineering, University of California at Los Angeles, CA 90095}

\maketitle

\begin{abstract}
A status updating system is considered in which a variable length code is used to transmit messages to a receiver over a noisy channel. The goal is to optimize the codewords lengths such that successfully-decoded messages are {\it timely}. That is, such that the {\it age-of-information} (AoI) at the receiver is minimized. A hybrid ARQ (HARQ) scheme is employed, in which variable-length incremental redundancy (IR) bits are added to the originally-transmitted codeword until decoding is successful. With each decoding attempt, a non-zero processing delay is incurred. The optimal codewords lengths are analytically derived utilizing a {\it sequential differential optimization} (SDO) framework. The framework is general in that it only requires knowledge of an analytical expression of the positive feedback (ACK) probability as a function of the codeword length.
\end{abstract}

\section{Introduction}
{\let\thefootnote\relax\footnote{{This research is supported by National Science Foundation (NSF) grant CCF-1955660. Any opinions, findings, and conclusions or recommendations expressed in this material are those of the author(s) and do not necessarily reflect views of the NSF.
}}}

Status updating over noisy communication channels calls for careful coding design such that the delivered status update messages are as timely as possible. Using an age-of-information (AoI) metric to assess timeliness, defined as the time elapsed since the latest successfully-decoded message has been generated, our goal in this paper is to provide an analytical framework to optimize codewords lengths for variable length codes used in delivering timely updates.

Most previous work on systems that seek to optimize codewords for AoI minimization, as in, e.g., \cite{najm-age-mg11-harq, sac-age-mg1-harq, simeone-age-finite-code, yates-age-erase-code, baknina-age-coding, feng-age-rateless-codes, najm-age-erasure-coding, javani-aoi-erasure}, have mainly focused on two distinct approaches, fixed redundancy (FR), in which the message is communicated with a single fixed-length transmission, and infinite incremental redundancy (IIR) schemes in which the transmission length is increased one symbol at a time until decoding is successful. Real systems often use a hybrid ARQ (HARQ) approach, as in, e.g., \cite{parag-age-coding, ceran-age-harq, arafa-aoi-coding, huang-estimation-harq-control}, in which the message length can be variable-length, but not at a granularity of a single symbol.  HARQ systems feature an initial transmission followed by subsequent transmissions (of possibly varying lengths) of incremental redundancy that are guided by feedback from the receiver to the transmitter. 


%



With no delay associated with decoding or requesting incremental redundancy, the pure IIR scheme is expected to provide a better AoI than the HARQ scheme that restricts the number of incremental redundancy transmissions.  However, most real systems include a nonzero processing delay $\beta$ corresponding to the time that it takes to decode the received codeword, transmit a negative acknowledgement (NACK) to the transmitter, and receive a subsequent incremental redundancy transmission.  For a large enough $\beta$, this overhead significantly increases the AoI of the IIR approach and makes the HARQ approach preferable.

Optimizing the HARQ approach requires determination of the length of the initial transmission and each subsequent transmission of incremental redundancy.   Sequential differential optimization (SDO) \cite{VakiliniaITW2014,Vakilinia2016,wong-sdo-conv} identifies a sequence of HARQ transmission lengths that optimizes throughput.  For a specified maximum number of feedback transmissions and a maximum probability that the decoder fails to produce a positive acknowledgement (ACK) even when all possible incremental redundancy has been received, SDO finds the transmission lengths that minimize average blocklength.  SDO requires a known probability distribution on the probability of ACK at each cumulative blocklength, but works equally well for the variety of distributions that arise from different variable-length codes operating on different channels  \cite{wong-sdo-conv,WangISIT2017,HeiderzadehISIT2018}. The original formulation of SDO minimizes the average blocklength for a fixed {\em maximum} number of feedback transmissions. The recent paper \cite{wesel-sdo-throughput}  re-frames the optimization problem using a Lagrangian approach to provide a closed-form expression for the optimal transmission lengths under a constraint on the {\em average} number of feedback transmissions.

{\it This paper extends the SDO approach to determine transmission lengths that explicitly optimize AoI.} Using AoI as the SDO objective function yields  different optimal transmission lengths than using throughput as the objective function as in \cite{wesel-sdo-throughput}, since the two objectives behave differently, see, e.g., \cite{yates_age_1}. 


One can differentiate between the works in \cite{najm-age-mg11-harq, sac-age-mg1-harq, simeone-age-finite-code, yates-age-erase-code, baknina-age-coding, feng-age-rateless-codes, najm-age-erasure-coding, javani-aoi-erasure, parag-age-coding, ceran-age-harq, arafa-aoi-coding, huang-estimation-harq-control} according to 1) whether status updates are exogenous or generated at will, depending on the ability to control transmission times; and 2) whether or not replacements are allowed, depending on the ability to let new updates replace the ones in service. Our work in this paper is categorized as a {\it generate-at-will HARQ scheme without replacement,} and is different from related works in that a nonzero processing delay $\beta$ is considered, and that the optimal set of codewords lengths that minimize the long-term average AoI is {\it analytically} derived.

Our case study for tail-biting convolutional codes shows that optimized HARQ beats optimized IIR and FR without replacement for all values of processing delay $\beta$.



\section{System Model and Problem Formulation} \label{sec_sys_mod}

\begin{figure}[t]
\center
\includegraphics[scale=.37]{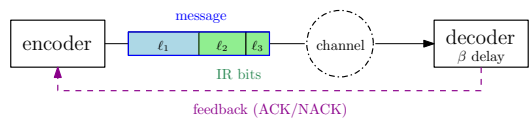}
\caption{Overview of the considered HARQ system model. In this example, $3$ transmissions are made before successful decoding, thereby requiring $\ell_2+\ell_3$ IR bits to be transmitted on top of the original $\ell_1$ bits. A processing delay of $3\beta$ time units is incurred in total ($\beta$ per decoding attempt).}
\label{fig_sys_mod}
\vspace{-.1in}
\end{figure}

We consider a transmitter-receiver pair communicating over a noisy memoryless channel. The transmitter generates $k$-bit measurements, at will, from a time-varying process. Measurements are time-stamped and sent to the receiver using $\ell_1$-bit codewords, $\ell_1\geq k$. We use the term {\it message} to denote a transmitted codeword. The receiver sends an ACK (a NACK) feedback following successful (unsuccessful) decoding attempts. Feedback messages are assumed to be free of errors, which is a mild assumption given the low information rate of the ACKs and NACKs. In addition to the time for message transmission, a fixed $\beta$ amount of time is consumed per decoding attempt, which includes the roundtrip time for sending feedback and processing it at the transmitter. We term $\beta$ the {\it processing delay}. A HARQ scheme is employed, in which IR bits are transmitted to help the receiver re-attempt decoding in case a NACK is fed back. IR lengths are denoted by $\{\ell_2,\ell_3,\dots,\ell_m\}$, where $m$ is the maximum number of transmission attempts per message. A system model overview is shown in Fig.~\ref{fig_sys_mod}.

Let us denote the cumulative blocklength by
\begin{align}
N_f\triangleq\sum_{i=1}^f\ell_i,\quad 1\leq f\leq m,
\end{align}
and let $P_{ACK}^{(N_f)}$ denote the probability of receiving an ACK while using a blocklength of $N_f$ bits. Clearly, such probability increases with $N_f$. The value of $N_m$ is  chosen to be large-enough that $P_{ACK}^{(N_m)}\approx1$, which depends on the specific code being used and the channel statistics.\footnote{We assume an ACK always corresponds to a successful (correct) decoding event. We ignore events in which an error bypasses the receiver undetected.} We note that $N_m$ is fixed, yet the value of $m$ is {\it not}; it is to be optimally-determined. Our SDO methodology, however, can be altered to work for fixed $N_m$ {\it and} $m$ (cf. Section~\ref{sec_conv_fix_m}).\footnote{Other cases, such as when $N_m$ is variable and $m$ is fixed, or when both are variable, are to be studied in future work.}

Let $\tau_i$ denote the $i$th {\it service time}: time consumed in transmitting the $i$th message. We consider a normalized setting in which sending a message using $N_f$ bits consumes $N_f$ time units. The channel is memoryless, and hence $\tau_i$'s are independent and identically distributed (i.i.d.) $\sim\tau$, which is approximately given by
\begin{align} \label{eq_tau}
\tau=\begin{cases}
N_1+\beta,\quad&\text{w.p. }P_{ACK}^{(N_1)} \\
N_f+f\beta,\quad&\text{w.p. }P_{ACK}^{(N_f)}-P_{ACK}^{(N_{f-1})},~f\geq2 
\end{cases}.
\end{align}
The above serves as a close approximation to $\tau$ under the reasonable assumption that receiving an ACK using $N_f$ bits implies receiving an ACK using $N_{f+1}$ bits as well. For instance, for $f=2$, one can write
\begin{align}
\mathbb{P}&\left(\tau=N_2+2\beta\right)=\mathbb{P}\left(\text{NACK at $N_1$, ACK at $N_2$}\right) \nonumber \\
&=\mathbb{P}\left(\text{ACK at $N_2$}\right)-\mathbb{P}\left(\text{ACK at $N_1$, ACK at $N_2$}\right) \nonumber \\
&=P_{ACK}^{(N_2)}-P_{ACK}^{(N_1)}+\mathbb{P}\left(\text{ACK at $N_1$, NACK at $N_2$}\right),
\end{align}
whence the last term is assumed having probability $\approx0$. Similar arguments can be followed for $f>2$.



Our goal is to design the blocklengths $\{N_f\}$ such that the long-term average AoI is minimized. The AoI at time $t$ is
\begin{align}
a(t)\triangleq t-u(t),
\end{align}
where $u(t)$ represents the time stamp of the latest successfully-decoded message. To minimize AoI, therefore, the transmitter should not acquire the $(i+1)$th measurement until the $i$th message is transmitted successfully, i.e., after (at least) $\tau_i$ time units starting from the transmission time of the $i$th message.

\begin{remark}
It is important to note that we focus on analyzing a HARQ scheme without replacement. Specifically, it might be better, AoI-wise, to drop the current message in transmission after a certain number of NACKs, and replace it by a new, fresher, one instead. This idea has been studied in, e.g., \cite{arafa-aoi-coding} for a system with fixed $m=2$. In this paper, we do not focus on systems that allow replacements. Instead, we aim at providing an analytical framework to design the blocklengths $\{N_f\}$ through a novel SDO approach discussed in Section~\ref{sec_sdo}.
\end{remark}

\begin{figure}[t]
\center
\includegraphics[scale=.32]{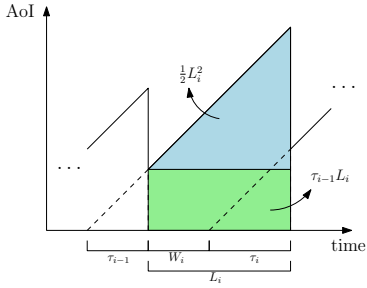}
\vspace{-.1in}
\caption{An example of how the AoI may evolve in the $i$th epoch.}
\label{fig_aoi_ex}
\vspace{-.1in}
\end{figure}

Let us denote by an epoch the time elapsed in between two successful transmissions. At the beginning of the $i$th epoch, the transmitter idly waits for $W_i$ time units before acquiring a new sample. Idle waiting can indeed minimize the average AoI as shown in various results of the literature, e.g., \cite{sun-age-mdp, arafa-aoi-estimate-ou}. In Fig.~\ref{fig_aoi_ex}, we show an example of how the AoI may evolve during the $i$th epoch. From the figure, one can see that the $i$th epoch length is given by
\begin{align} \label{eq_L_i}
L_i=W_i+\tau_i,
\end{align}
and the corresponding area under the AoI curve is
\begin{align} \label{eq_Q_i}
Q_i=\tau_{i-1}L_i+\frac{1}{2}L_i^2.
\end{align}
The sequence $\{W_i\}$ denotes a {\it waiting policy}. Our goal is to find the optimal blocklenghts and waiting policy that minimize the long-term average AoI given by
\begin{align} \label{eq_long_term_aoi}
\limsup_{j\rightarrow\infty} \frac{\sum_{i=1}^j\mathbb{E}\left[Q_i\right]}{\sum_{i=1}^j\mathbb{E}\left[L_i\right]}.
\end{align}

Since $\tau_i$'s are i.i.d., one can then conclude using the results in \cite{sun-age-mdp} that the optimal waiting policy has a threshold structure, in which
\begin{align} \label{eq_W_i}
W_i=\left[\gamma-\tau_{i-1}\right]^+,
\end{align}
where $\gamma\geq0$ is some threshold, and $[\cdot]^+\triangleq\max(\cdot,0)$. This induces a {\it stationary} distribution $L_i\sim L$ and $Q_i\sim Q$ for all epochs, and thereby reduces the focus to a typical epoch through removing the summations in the numerator and denominator of (\ref{eq_long_term_aoi}). Let us define $\overline{\tau}$ as the starting AoI of such an epoch. This allows us to write
\begin{align}
\mathbb{E}\left[L\right]\!=&\mathbb{E}\left[\left[\gamma-\overline{\tau}\right]^+\right]+\mathbb{E}\left[\tau\right], \label{eq_L} \\
\mathbb{E}\left[Q\right]\!=&\mathbb{E}\!\left[\overline{\tau}\left[\gamma\!-\!\overline{\tau}\right]^+\right]\!+\!\left(\mathbb{E}\!\left[\tau\right]\right)^2\!+\!\frac{1}{2}\mathbb{E}\!\left[\left(\left[\gamma-\overline{\tau}\right]^+\!+\!\tau\right)^2\right]. \label{eq_Q}
\end{align}
Our optimization problem is therefore given by
\begin{align} \label{opt_main_wait}
\min_{\{N_f\},~\gamma\geq0} \quad &\frac{\mathbb{E}\left[Q\right]}{\mathbb{E}\left[L\right]} \nonumber \\
\mbox{s.t.}~~~\quad&N_f>N_{f-1},~N_f\in\mathbb{Z}_{++},~\forall f 
\end{align}
with $N_0\triangleq k$; $\mathbb{E}\left[L\right]$ and $\mathbb{E}\left[Q\right]$ given by (\ref{eq_L}) and (\ref{eq_Q}), respectively; and $\overline{\tau}$ and $\tau$ i.i.d. as in (\ref{eq_tau}).

One can possibly follow a decomposition approach to solve problem (\ref{opt_main_wait}) by fixing the threshold $\gamma$ and solving for the blocklengths $\{N_f\}$ in terms of $\gamma$, and then finding the optimal threshold afterwards. We realize, however, that such approach would not yield a clear analytical solution for the blocklengths, {\it which is one fundamental goal for this paper.} Thereby, in Section~\ref{sec_sdo}, we focus on problem (\ref{opt_main_wait}) in the special case of a {\it zero-wait policy,} i.e., when $\gamma=0$, and present a novel SDO framework to find the optimal blocklengths. After that, in Section~\ref{sec_wait}, we discuss how to find the threshold based on the SDO solution (which may be suboptimal). Under a zero-wait policy, the objective function of problem (\ref{opt_main_wait}) is simplified to
\begin{align} \label{eq_rho_0}
\rho_0\triangleq\mathbb{E}\left[\tau\right]+\frac{\mathbb{E}\left[\tau^2\right]}{2\mathbb{E}\left[\tau\right]}.
\end{align}

\section{The SDO Approach} \label{sec_sdo}

In this section, we solve problem (\ref{opt_main_wait}) for $\gamma=0$. The SDO approach basically solves for all the blocklengths sequentially in terms of $N_1$. A one-dimensional search is then followed to find the optimal $N_1^*$, and subsequently all the other blocklengths. Such approach, however, will {\it not} work if we optimize $\rho_0$ in its current fractional form. The reason, for instance, is that the partial derivative of $\rho_0$ with respect to $N_1$ is a function of all the blocklengths, while it should only be a function of $N_1$ and $N_2$ so that the optimal $N_2$ can be completely characterized in terms of $N_1$.

In fact, as we will show, the SDO approach {\it will} work if $\rho_0$ is represented in an equivalent yet non-fractional way. Towards that end, we follow a Dinkelbach-like approach \cite{dinkelbach-fractional-prog}, and introduce the following auxiliary problem for fixed $\lambda\geq0$:
\begin{align} \label{opt_aux}
p(\lambda)\triangleq\min_{\{N_f\}}\quad&(1-\lambda)\mathbb{E}\left[\tau\right]+\frac{1}{2}\mathbb{E}\left[\tau^2\right] \nonumber \\
\mbox{s.t.}\quad&N_f>N_{f-1},~N_f\in\mathbb{Z}_{++},~\forall f.
\end{align}
Let $\rho_0^*$ denote the optimal long-term average AoI in (\ref{eq_rho_0}). We now have the following result:

\begin{lemma} \label{thm_lmda}
Let $\{N_f^{\lambda}\}$ denote the solution of problem (\ref{opt_aux}), and $\tau_\lambda$ be the corresponding service time. It then holds that 
\begin{align}
\rho_0^*=p\left(\lambda^*\right)+\lambda^*,
\end{align}
where $\lambda^*\triangleq\arg\min\{p(\lambda)+\lambda:~p(\lambda)=\mathbb{E}\left[\tau_{\lambda}\right]\}$. 
\end{lemma}
\begin{Proof}
First, it is direct to see that $p(\lambda)=\mathbb{E}\left[\tau_\lambda\right]\iff\lambda=\frac{\mathbb{E}\left[\tau_\lambda^2\right]}{2\mathbb{E}\left[\tau_\lambda\right]}$,
and that at such case $\rho_0$ would be equal to $p(\lambda)+\lambda$. It therefore follows that $\rho_0^*$ is given by minimizing the expression $p(\lambda)+\lambda$ over all values of $\lambda$ that satisfy $p(\lambda)=\mathbb{E}\left[\tau_\lambda\right]$. Next, one can show that $p(\lambda)$ is decreasing in $\lambda$. In particular, there exists some $\lambda_{\max}$ such that $p(\lambda_{\max})<0$. This shows that the set $\{\lambda:~p(\lambda)=\mathbb{E}\left[\tau_{\lambda}\right]\}$ is non-empty and $\lambda^*$ exists.
\end{Proof}

Lemma~\ref{thm_lmda} shows that one can find the optimal long-term average AoI in (\ref{eq_rho_0}) by focusing on solving problem (\ref{opt_aux}) at a specific $\lambda^*$. The value of $\lambda^*$ can be found via, e.g., a one-dimensional search over the interval $\left[0,\lambda_{\max}\right]$, where $\lambda_{\max}$ is a large-enough value of $\lambda$ such that $p(\lambda_{\max})<0$. We observe that for the case of the convolutional codes studied in Section~\ref{sec_conv}, such $\lambda^*$ is also unique (cf. Fig.~\ref{fig_p_lmda_tau_lmda_beta10}). 

Given this auxiliary result, we now discuss how to use SDO to find the optimal codewords lengths for fixed $\lambda$ by solving problem (\ref{opt_aux}). First, let us relax the problem by ignoring the integer constraints on the blocklengths and solving for real values of $\{N_f\}$. Imposing the integer constraints back on the acquired solutions can be handled, e.g., via the dithering approach proposed in \cite[Section IV-B]{wesel-sdo-throughput}. In our work, we follow a rounding approach instead to project the optimal blocklengths onto $\mathbb{Z}_{++}$, yet we do so {\it simultaneously} after solving for all of them. We observe that such rounding approach has a negligible effect on optimality especially for relatively large blocklengths, as discussed in Section~\ref{sec_conv}.

Next we elaborate on the partial derivatives of the first and second moments of $\tau$ with respect to the blocklenghts $\{N_f\}$. Using (\ref{eq_tau}), the first moment is given by
\begin{align}
\mathbb{E}\left[\tau\right]=&\left(N_1+\beta\right)P_{ACK}^{(N_1)} \nonumber \\
&+\sum_{f=2}^{m-1}\left(N_f+f\beta\right)\left(P_{ACK}^{(N_f)}-P_{ACK}^{(N_{f-1})} \right) \nonumber \\
&+\left(N_m+m\beta\right)\left(1-P_{ACK}^{(N_{m-1})} \right),
\end{align}
whose partial derivatives are given by
\begin{align}
\frac{\partial\mathbb{E}\left[\tau\right]}{\partial N_1}=&P_{ACK}^{(N_1)}+\left(N_1+\beta-\left(N_2+2\beta\right)\right)P_{ACK}^{\prime(N_1)}, \\
\frac{\partial\mathbb{E}\left[\tau\right]}{\partial N_f}=&P_{ACK}^{(N_f)}-P_{ACK}^{(N_{f-1})} \nonumber \\
&+\left(N_f+f\beta-\left(N_{f+1}+(f+1)\beta\right)\right)P_{ACK}^{\prime(N_f)}, 
\end{align}
for $2\leq f\leq m-1$, where $P_{ACK}^{\prime(N_f)}$ denotes the derivative $\frac{dP_{ACK}^{(N_f)}}{dN_f}$. Similarly, the second moment is expressed as

\begin{align}
\mathbb{E}\left[\tau^2\right]=&\left(N_1+\beta\right)^2P_{ACK}^{(N_1)} \nonumber \\
&+\sum_{f=2}^{m-1}\left(N_f+f\beta\right)^2\left(P_{ACK}^{(N_f)}-P_{ACK}^{(N_{f-1})} \right) \nonumber \\
&+\left(N_m+m\beta\right)^2\left(1-P_{ACK}^{(N_{m-1})} \right),
\end{align}
whose partial derivatives are given by
\begin{align}
\frac{\partial\mathbb{E}\left[\tau^2\right]}{\partial N_1}=&2\left(N_1+\beta\right)P_{ACK}^{(N_1)} \nonumber \\
&+\left(\left(N_1+\beta\right)^2-\left(N_2+2\beta\right)^2\right)P_{ACK}^{\prime(N_1)}, \\
\frac{\partial\mathbb{E}\left[\tau^2\right]}{\partial N_f}=&2\left(N_f+f\beta\right)\left(P_{ACK}^{(N_f)}-P_{ACK}^{(N_{f-1})}\right) \nonumber \\
&\hspace{-.5in}+\left(\left(N_f+f\beta\right)^2-\left(N_{f+1}+(f+1)\beta\right)^2\right)P_{ACK}^{\prime(N_f)}, 
\end{align}
for $2\leq f\leq m-1$.

Now let us take the partial derivative of the objective function of problem (\ref{opt_aux}) with respect to $N_1$ and equate it to 0. Using the above, after some algebra we get that
\begin{align} \label{eq_n2_n1_quad}
\left(N_2+2\beta\right)^2+2(1-\lambda)\left(N_2+2\beta\right)-c\left(N_1,\lambda\right)=0
\end{align}
must hold, where
\begin{align}
c\left(N_1,\lambda\right)\triangleq&2(1-\lambda)\left(\frac{P_{ACK}^{(N_1)}}{P_{ACK}^{\prime(N_1)}}+\left(N_1+\beta\right)\right) \nonumber \\
&+2\left(N_1+\beta\right)\left(\frac{P_{ACK}^{(N_1)}}{P_{ACK}^{\prime(N_1)}}+\frac{\left(N_1+\beta\right)}{2}\right).
\end{align}
Now let us fix the value of $N_1$ ($\geq k$). If the discriminant of the quadratic equation in (\ref{eq_n2_n1_quad}), i.e., if
\begin{align}
\left(1-\lambda\right)^2+c\left(N_1,\lambda\right)
\end{align}
is negative, then there do not exist any real solutions for $N_2$ that solve (\ref{eq_n2_n1_quad}). This means that the fixed value of $N_1$ is {\it not optimal,} and has to change. On the other hand, if the above discriminant is non-negative, then one can get the following two solutions for $N_2$:
\begin{align} \label{eq_n2_n1}
N_2=-\left(1-\lambda\right)\pm\sqrt{\left(1-\lambda\right)^2+c\left(N_1,\lambda\right)}-2\beta.
\end{align}
Similarly, one can show that taking the partial derivative of the objective function of problem (\ref{opt_aux}) with respect to $N_f$, $2\leq f\leq m-1$, and equating it to $0$ results in a quadratic equation to solve for $N_{f+1}$ in terms of $N_f$ and $N_{f-1}$. The two solutions of such equation are given by
\begin{align} \label{eq_nf+1_nf_nf-1}
N_{f+1}\!=\!&-\left(1\!-\!\lambda\right)\pm\sqrt{\left(1\!-\!\lambda\right)^2\!+\!c\left(N_f,N_{f-1},\lambda\right)}-(f\!+\!1)\beta,\end{align}
where
\begin{align}
&\hspace{-.25in}c(N_f,N_{f-1},\lambda) \nonumber \\
\triangleq&2(1\!-\!\lambda)\!\left(\frac{P_{ACK}^{(N_f)}\!-\!P_{ACK}^{(N_{f-1})}}{P_{ACK}^{\prime(N_f)}}\!+\!\left(N_f\!+\!f\beta\right)\right) \nonumber \\
&+\!2\left(N_f\!+\!f\beta\right)\!\left(\frac{P_{ACK}^{(N_f)}\!-\!P_{ACK}^{(N_{f-1})}}{P_{ACK}^{\prime(N_f)}}\!+\!\frac{\left(N_f\!+\!f\beta\right)^2}{2}\right),
\end{align}
provided that the discriminant below is non-negative:
\begin{align}
\left(1-\lambda\right)^2+c\left(N_f,N_{f-1},\lambda\right).
\end{align}

Therefore, using (\ref{eq_n2_n1}) and (\ref{eq_nf+1_nf_nf-1}), one can characterize optimal solutions for $\{N_2,N_3,\dots,N_{m-1}\}$ in terms of $N_1$. These sequential solutions would eventually stop if $N_{f^*+1}$ surpasses $N_m$, for some $f^*$, at which point one may truncate the excess IR bits and set $N_{f^*+1}=N_m$.

Now for the solutions to be meaningful, we need to make sure that the obtained blocklengths are monotonically increasing. In most scenarios, such as in the one discussed in Section~\ref{sec_conv}, this would automatically cross-out the smaller solutions in (\ref{eq_n2_n1}) and (\ref{eq_nf+1_nf_nf-1}), especially for large values of $f$.

For $2\leq f\leq m-1$, in case both solutions obtained for $N_f$ are smaller than $N_{f-1}$, or in case the discriminant of the quadratic equation to solve for $N_{f+1}$ is negative, then the whole solution sequence leading to such $N_f$ is {\it rejected.} If it so happens that all solution sequences are rejected, then the fixed value of $N_1$ is not optimal, and has to change. As noted in Section~\ref{sec_conv}, we observe that for large values of $\beta$, one needs to initiate SDO with a relatively large value of $N_1$ to get meaningful (unrejected) solution sequences. Finally, in case two or more solution sequences are obtained, we pick the one that yields a smaller objective function of problem (\ref{opt_aux}).

We now summarize the SDO approach used to characterize the optimal long-term average AoI $\rho_0^*$. For a given $\lambda$, we first fix $N_1$ and sequentially solve for $\{N_2,N_3,\dots,N_{m-1}\}$ using equations (\ref{eq_n2_n1}) and (\ref{eq_nf+1_nf_nf-1}). We then find the best $N_1$, which gives $p(\lambda)$. Finally, the optimal $\lambda^*$ is found as discussed in Lemma~\ref{thm_lmda}, which gives $\rho_0^*=p(\lambda^*)+\lambda^*$. 

%


\section{Waiting Policy} \label{sec_wait}

We now consider optimizing the waiting policy by going back to problem (\ref{opt_main_wait}). As discussed towards the end of Section~\ref{sec_sys_mod}, jointly optimizing the waiting threshold $\gamma$ and the blocklenghts $\{N_f\}$ would not directly yield a sequential solution as done in the previous section. We instead follow a potentially-suboptimal approach in which we first find the optimal blocklengths via SDO for a zero-wait policy, {\it then} we optimize the waiting threshold based on that. Therefore, in this section we assume that we already have a set of blocklengths $\{N_f\}$, with a corresponding service time random variable $\tau$. Now the task of finding the optimal $\gamma^*$ can be accomplished by the techniques introduced in \cite{sun-age-mdp}. In what follows, we reiterate the procedure of finding $\gamma^*$ according to our own notation, and approach it slightly differently, for completeness.

To analytically determine the optimal threshold $\gamma^*$, one can leverage (the original) Dinkelbach's approach \cite{dinkelbach-fractional-prog} for some fixed $\eta\geq0$ and define
\begin{align}
q(\eta)\triangleq \min_{\gamma\geq0} \mathbb{E}\left[Q\right]-\eta\mathbb{E}\left[L\right],
\end{align}
with  $\mathbb{E}\left[L\right]$ and $\mathbb{E}\left[Q\right]$ given by (\ref{eq_L}) and (\ref{eq_Q}), respectively. Next, one can show that the following holds:
\begin{align}
\frac{d\mathbb{E}\left[Q\right]}{d\gamma}\!=\!\left(\gamma+\mathbb{E}\left[\tau\right]\right)\mathbb{P}\left(\tau\leq\gamma\right),~\frac{d\mathbb{E}\left[L\right]}{d\gamma}\!=\!\mathbb{P}\left(\tau\leq\gamma\right).
\end{align}
Therefore, after setting $\frac{d\left(\mathbb{E}\left[Q\right]-\eta\mathbb{E}\left[L\right]\right)}{d\gamma}=0$, the optimal threshold will be given by
\begin{align} \label{eq_wait_threshold}
\gamma^*=\eta^*-\mathbb{E}\left[\tau\right],
\end{align}
where $\eta^*$ is the unique solution of $q(\eta^*)=0$, which can be found via, e.g., a bisection search \cite{dinkelbach-fractional-prog}.

We note that $\gamma^*>0$, and is therefore a meaningful threshold. This can be seen by observing that
\begin{align}
q\left(\mathbb{E}\!\left[\tau\right]\right)\!=\!\mathbb{E}\!\left[\overline{\tau}\left[\gamma\!-\!\overline{\tau}\right]^+\right]\!+\!\frac{1}{2}\mathbb{E}\!\left[\!\left(\left[\gamma\!-\!\overline{\tau}\right]^+\right)^2\right]\!+\!\frac{1}{2}\mathbb{E}\!\left[\tau^2\right],
\end{align}
which is strictly positive. Since $q(\eta)$ is decreasing \cite{dinkelbach-fractional-prog}, we must have $\eta^*>\mathbb{E}\left[\tau\right]$ in order for $q(\eta^*)=0$ to hold.

\section{Case Study: Convolutional Codes} \label{sec_conv}

We apply the above analysis to the case of tail-biting convolutional codes over additive white Gaussian noise (AWGN) channels. As shown in \cite{wong-sdo-conv} for binary inputs with a signal-to-noise ratio (SNR) of 2 dB, the Gaussian distribution closely-approximates the ACK probability as follows:
\begin{align} \label{eq_p-ack_convl}
P_{ACK}^{(N_f)}\approx Q\left(\frac{k/N_f-0.5666}{0.0573}\right),
\end{align}
where $Q(x)\triangleq\frac{1}{\sqrt{2\pi}}\int_x^\infty e^{\frac{-u^2}{2}}du$ is the Q-function. We set the measurement length to $k=64$ bits and $N_m=192$ bits. Our results are in the context of the model in (\ref{eq_tau}) and (\ref{eq_p-ack_convl}).

\subsection{Verifying Lemma~\ref{thm_lmda}}

\begin{figure}[t]
\center
\includegraphics[scale=.35]{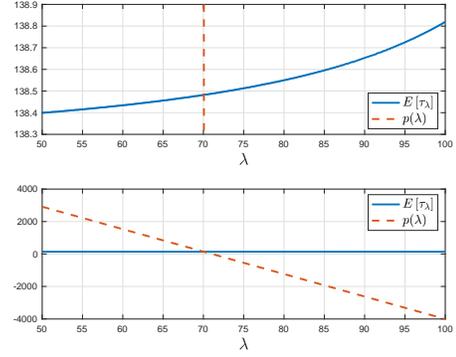}
\vspace{-.1in}
\caption{$p(\lambda)$ and the optimal average service time $\mathbb{E}\left[\tau_{\lambda}\right]$ vs. $\lambda$, with $\beta=10$ time units. Top plot is a zoomed-in version of bottom plot. There exists a unique $\lambda^*\approx70$ such that $p\left(\lambda^*\right)=\mathbb{E}\left[\tau_{\lambda^*}\right]$, at which $p\left(\lambda^*\right)\approx138$.}
\label{fig_p_lmda_tau_lmda_beta10}
\vspace{-.1in}
\end{figure}

We first verify the results of Lemma~\ref{thm_lmda}. For a system with $\beta=10$ time units, we plot both $\mathbb{E}\left[\tau_{\lambda}\right]$ and $p(\lambda)$ versus $\lambda$ in Fig.~\ref{fig_p_lmda_tau_lmda_beta10}. We see that $\mathbb{E}\left[\tau_{\lambda}\right]$ is increasing with $\lambda$. This makes the set $\{\lambda:p(\lambda)=\mathbb{E}\left[\tau_{\lambda}\right]\}$ basically a {\it singleton,} which further facilitates evaluating $\lambda^*$ through a bisection search over $[0,\lambda_{\max}]$. We note that such case holds for all values of $\beta$.

\begin{figure}[t]
\center
\includegraphics[scale=.35]{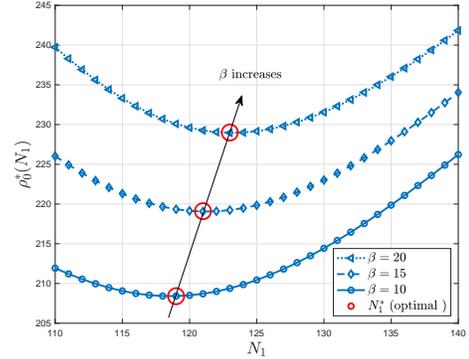}
\vspace{-.1in}
\caption{Optimal long-term average AoI as a function of $N_1$, with different $\beta$'s. The optimal $N_1^*$ is denoted by red circles. For $\beta=10$, the optimal $N_1^*=119$ bits, with $\rho_0^*(119)\approx208$ time units.}
\label{fig_rho_N1_beta10}
\vspace{-.1in}
\end{figure}

Next, we show how the optimal long-term average AoI behaves as a function of $N_1$. That is, we solve for $\rho_0^*(N_1)$ as opposed to $\rho_0^*$. We do so via slightly modifying the SDO approach. Specifically, now that $N_1$ is fixed, we substitute in (\ref{eq_rho_0}) to get a relatively new metric $\rho_0(N_1)$ to be optimized by choosing $\{N_2,N_3,N_4,\dots,N_{m-1}\}$. For that, we follow the same SDO approach discussed in Section~\ref{sec_sdo}, yet after replacing $N_1$ with $N_2$. The result is shown in Fig.~\ref{fig_rho_N1_beta10} for $\beta\in\{10,15,20\}$. We see that the optimal $N_1^*$ that minimizes $\rho_0^*(N_1)$ is relatively mid-range and, intuitively, increases with $\beta$. Combining the results of Fig.~\ref{fig_p_lmda_tau_lmda_beta10} and Fig.~\ref{fig_rho_N1_beta10}, we observe that at $\beta=10$, $p\left(\lambda^*\right)+\lambda^*=\rho_0^*\left(N_1^*\right)$, as asserted in Lemma~\ref{thm_lmda}.

\subsection{A Methodology for fixed $m$} \label{sec_conv_fix_m}

\begin{figure}[t]
\center
\includegraphics[scale=.35]{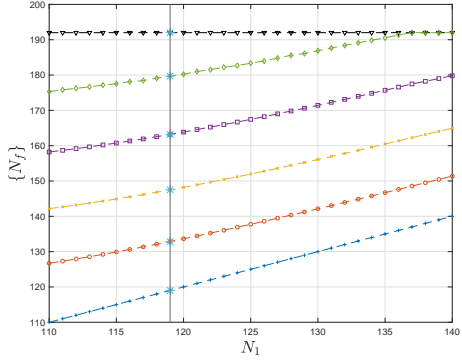}
\vspace{-.1in}
\caption{Optimal IR lengths vs. $N_1$ using SDO, with $\beta=10$ time units. The optimal set of blocklengths is at $N_1^*=119$ and are denoted by $\ast$.}
\label{fig_IR_N1_beta10}
\vspace{-.15in}
\end{figure}

In Fig.~\ref{fig_IR_N1_beta10}, we show how the optimal blocklengths vary with $N_1$ for $\beta=10$. We see that as $N_1$ increases, the set of blocklengths becomes sparser, i.e., fewer number of IR transmissions leads to reaching $N_m$. This figure, together with Fig.~\ref{fig_rho_N1_beta10} can be used to solve the problem with {\it fixed number of transmissions per message $m$,} which may be relevant in some practical systems. For instance, at $N_1^*=119$ we have $m=6$ transmissions. If we have a constraint of only $m=5$, then we would have to use $N_1\geq137$ according to Fig.~\ref{fig_IR_N1_beta10}. We would then examine Fig.~\ref{fig_rho_N1_beta10} to conclude that $N_1=137$ is the optimal choice in this case since it attains the smallest AoI for $\beta=10$ when compared to higher values of $N_1$.


\subsection{Comparison to Baseline Schemes: IIR and FR}

We compare the proposed HARQ scheme with other baseline schemes. The first is IIR, in which incremental bits are added {\it one-by-one} until success. This is a special case of HARQ in which $N_{f+1}=N_f+1,~\forall f$ (presuming that $m$ can be arbitrarily large). The second baseline scheme is FR, for which we consider two subcases: with and without replacement. FR without replacement is basically using a fixed $N_1$ to transmit each message, with repetition in case of failures. This makes the service time given by $(N_1+\beta)M$, where $M$ is a geometric random variable with parameter $P_{ACK}^{(N_1)}$. FR with replacement is strictly better than FR without replacement in the sense it uses fresh measurements after failures. This makes the epoch length also given by $(N_1+\beta)M$, yet the service time is {\it fixed} at $N_1+\beta$. For IIR and FR without replacement, one can jointly optimize $N_1$ and the optimal waiting threshold in (\ref{eq_wait_threshold}).\footnote{Different from HARQ, this joint optimization can be optimally solved.} For FR with replacement, a zero-wait policy is optimal, see \cite[Theorem 2]{arafa-aoi-estimate-ou}, and the long-term average AoI can be shown to be equal to $\left(N_1+\beta\right)\left(1/P_{ACK}^{(N_1)}+1/2\right)$.

\begin{figure}[t]
\center
\includegraphics[scale=.35]{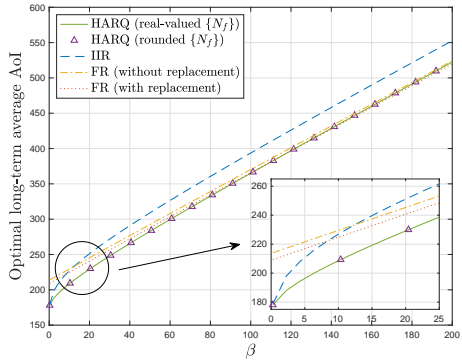}
\vspace{-.1in}
\caption{Proposed HARQ and baselines (IIR, and FR with and without replacement) vs. $\beta$. Triangles denote rounded (integer) blocklengths.}
\label{fig_aoi_IIR_FR_HARQ_beta}
\vspace{-.15in}
\end{figure}

Fig.~\ref{fig_aoi_IIR_FR_HARQ_beta} shows the optimal long-term average AoI for the proposed HARQ scheme and baseline IIR and FR as a function of $\beta$. We also plot the AoI achieved by HARQ after rounding the blocklengths to their nearest integer values; we see that the performance is almost identical after rounding as noted in Section~\ref{sec_sdo}. The HARQ scheme outperforms IIR and FR without replacement for all values of $\beta$. It outperforms FR with replacement for $\beta\lesssim120$. For $\beta\gtrsim120$ HARQ AoI is slightly above FR with replacement since optional replacement is not included in the current analysis of HARQ.

\section{Conclusion}

An SDO-based analytical framework has been developed to produce AoI-minimal HARQ transmission lengths. Different from almost all of the AoI-related literature on coding design, a nonzero processing delay is considered in our system, which includes the time to decode a message, send feedback and initiate the transmission of IR bits if needed. The optimized HARQ scheme beats multiple baselines such as IIR and FR.

Future work includes developing an SDO-based framework for HARQ in systems that allow message replacement.

\end{document}